\documentstyle[twocolumn,epsf,aps,prl]{revtex} 
\begin{document} %****
\title{ 
%Novel Natural Boundaries for Quantum Phase Transitions 
%Reconciling Two Estimates of the Magnetic Correlation Length
%in the Hole-doped Cuprates}
Magnetic Quantum Critical Point in YBa$_2$Cu$_3$O$_{7-\delta}$?\\
Resolving the Correlation Length Controversy}
\author{R.S. Markiewicz}
\address{
Physics Department, Northeastern University, Boston MA 02115, USA}
\maketitle
\begin{abstract}
A mode coupling calculation which previously explained the Mott gap
collapse induced in cuprates by {\it electron} doping is applied to the
analogous problem of {\it hole} doping.  A plateau in the $\vec q$-space
susceptibility is found to inhibit the rate of susceptibility divergence
as $T\rightarrow 0$.  This effect could reconcile neutron and NMR
measurements of the correlation length in this regime, and clarify the
issue of proximity to a quantum critical point.
\end{abstract}
%\pacs{PACS numbers~:~~74.20.Mn, 74.72.-h, 71.45.Lr, 74.50.+r }

%****
\narrowtext
%****
%%%%New Start
There is currently a controversy as to the magnetic correlation length
$\xi$ in hole doped cuprates, with neutron scattering\cite{Bou} finding a
T-independent value while NMR studies\cite{MMP,OBAM} suggest that $\xi$
increases as T decreases.  While the numerical values are not greatly
dissimilar, the temperature dependence is important, since a
weakly-diverging $\xi$ -- even if ultimately cut off by superconductivity
-- could provide evidence for proximity to a quantum phase transition (QPT),
as has been suggested to occur slightly above optimal doping\cite{Tal1}.
Here, a resolution of this problem is suggested, which is consistent with
the theoretical model of the QPT.

Recent evidence for a QPT in electron-doped cuprates\cite{nparm,Greene}
has generated a lot of theoretical interest\cite{KLBM,MKII,KuR,SeMSTr}. 
It is found that at a critical electron doping, there is a QPT where (a)
the Mott gap collapses, (b) the T=0 N\'eel transition terminates, and (c)
the Fermi surface crosses over from small pockets to a large barrel.  The
simplest mean field theory\cite{KLBM} can reproduce the transition, if a
Kanamori-like\cite{Kana} doping dependence of the Hubbard $U$ is assumed.
The results have been confirmed by a number of
calculations\cite{MKII,KuR,SeMSTr} which incorporate interactions with
spin waves, the main difference being that the mean-field gap and Neel
transition become a pseudogap and crossover temperature $T^*$.  [Secondary
interactions can generate a finite Neel temperature $T_N<<T^*$.]

Extension of the model to the {\it hole-doping} case is problematic. 
The model predicts a nearly electron-hole symmetric QPT, with pseudogap
collapse.  In fact, the observed pseudogap does follow 
the predicted doping dependence\cite{MKII}, terminating in a QPT\cite{Tal1} 
which is approximately electron-hole symmetric.  This symmetry can be seen
in the phase diagrams\cite{ADa} of electron-doped Nd$_{2-x}$Ce$_x$CuO$_{4
\pm\delta}$ (NCCO) and hole-doped La$_{2-x}$Sr$_x$CuO$_4$ (LSCO).  Signatures 
of strong local magnetic couplings disappear near the dopings of optimal
superconducting $T_c$; the corresponding dopings are comparable, and the
$T_c$ values themselves do not greatly differ.  Near optimal doping the
normal-state resistivity is linear in temperature\cite{FoGr}, suggestive
of a QPT.  Moreover, it was recently suggested\cite{MZG} that this QPT
involves a crossover to the large Fermi surface.

However, the third element of the transition, the $T=0$ N\'eel transition,
appears to be absent: the anticipated correlation length $\xi$ divergence
as $T\rightarrow 0$ is cut off at a surprisingly low doping in
neutron measurements\cite{Bou}.  
%This contradictory result could be due to frustration associated
%with competing order parameters (possibly related to stripe physics). 
On the other hand, the T-dependent $\xi$ deduced from NMR could suggest
that a {\it weak} $T\rightarrow 0$ Neel transition is still present.  The
questions then are two: can the NMR and neutron results be reconciled, and
why is the Neel transition so much weaker in hole-doped cuprates?  Both
questions can be answered by a self-consistent renormalization (SCR)
theory\cite{MKII,Mor}.

Within the SCR approach, the zero-temperature antiferromagnetic (AFM)
transition is controlled by a renormalized Stoner criterion\cite{MKII},
$U\chi_{0Q}=\eta$, with $U$ the Hubbard $U$, $\eta > 1$ a quantum
correction, and $\chi_{0Q}=\chi_0(\vec Q,\omega =0)$ the bare magnetic
susceptibility at $\vec Q=(\pi ,\pi)$,
\begin{equation}
\chi_{0Q}=\sum_{\vec k}{f(E_{\vec k})-f(E_{\vec k+\vec Q})\over
E_{\vec k}-E_{\vec k+\vec Q}},
\label{eq:1}
\end{equation}
where $f$ is the Fermi function, and $E$ the electronic dispersion, taken
to be of tight-binding form
\begin{equation}                                                                
E_{\vec k}=-2t(c_x+c_y)-4t'c_xc_y,
\label{eq:2} 
\end{equation}                                                                 
with $c_i=\cos{k_ia}$ and $a$ is the lattice constant.  At finite
temperatures, there is a Mott-like pseudogap in the density of states,
which first appears at a temperature close to the RPA N\'eel temperature
(for $U^*=U/\eta$) -- in particular, the pseudogap collapses to $T=0$ at
the zero temperature N\'eel transition.  While the doping at which this
QPT occurs depends sensitively on $|t'|$, the QPT falls at approximately the
same doping for both electron and hole doping.

\begin{figure}
\leavevmode
   \epsfxsize=0.33\textwidth\epsfbox{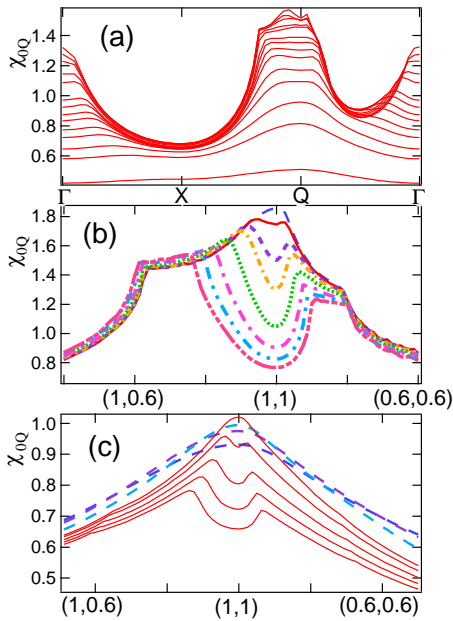}
\vskip0.5cm
\caption{(a) Susceptibility $\chi_{0\vec q}$ for several hole dopings in
the hot spot regime, at the mean-field N\'eel tempreature $T_N$.  From
lowest to highest curve, dopings $x$ [and $T_N$] are 0 [4700], 0.05
[2200], 0.078 [1520], 0.109 [961], 0.127 [701], 0.147 [475], 0.160 [369],
0.175 [278], 0.188 [222], 0.214 [133], 0.225 [105], 0.236 [70], 0.241
[55], and 0.246 [40K].
(b) Susceptibility $\chi_{0\vec q}$ for several hole dopings $x\ge x_H$
near $\vec Q$, at $T=10K$. From bottom to top (at $\vec Q$): $x$ = 0.300,
0.286, 0.274, 0.261, 0.253, 0.250, 0.247, and 0.246.
(c) Susceptibility $\chi_{0\vec q}$ for several electron dopings near
the C-point: for dashed lines from bottom to top: x [$T_N$] = -0.147
[1661], -0.171 [1304], and -0.196 [853K]. Solid lines, from top to bottom:
$x$ = -0.213 [C-point, $T_N=199K$], and, beyond C-point, $T=100K$, $x$ = 
-0.223, -0.246, -0.269, and -0.293.}
\label{fig:1}
\end{figure}

It is here suggested that the striking differences between electron and
hole doped cuprates can be understood from the Stoner criterion,
specifically, the doping dependence of $\chi_{0q}$ for $q$ near $Q$,
Fig.~\ref{fig:1}.  For hole doping, the susceptibility has the form of a
(diamond-shaped) plateau centered on $Q$; for electron-doping the plateau
width has shrunk nearly to zero.  Briefly, as $T$ decreases, the Stoner
enhancement arises {\it only very near $q=Q$ for electron-doped} systems;
as the peak height diverges the width (inverse correlation length) shrinks
to zero, and the area under the curve remains small, leading to a strong
transition, $\xi$ diverging exponentially in $1/T$.  For hole-doping, the
Stoner enhancement is spread over the full plateau.  This has two effects:
first, the susceptibility width is pinned at the plateau width, and is
insensitive to the $\xi$ divergence; second, the large area over which the
susceptibility is growing can lead to a sum-rule saturation and a very
sluggish divergence, $\xi\sim 1/\sqrt{T}$.  The detailed calculations
below confirm this scenario, and show that the $\xi$ measured from NMR has
the expected $T$-dependence.  Thus, the SCR model both explains the
seemingly contradictory neutron and NMR data, and simultaneously explains
the electron-hole asymmetry.

The susceptibility plateau is controlled by {\it hot spot}
physics, where a hot spot is defined as a point on the Fermi surface (FS)
which is separated from another FS point by exactly the antiferromagnetic
vector $\vec Q$.  For any $t'\ne 0$, the hot spots exist in a finite range
of doping about half filling; the critical dopings which terminate the hot
spot regime are called $x_C$ (electron doping) and $x_H$ (hole doping) --
the latter coinciding with the van Hove singularity (VHS).  Between these
two dopings, the susceptibility is large\cite{AAA,BCT,Tdep}, but outside
this doping range the susceptibility falls rapidly, so for a wide range of
couplings, $U$, the Stoner criterion will hold when hot spots are
present, but will fail when the hot spots disappear.  The approximate
electron-hole symmetry of the hot-spot regime leads to a corresponding
symmetry of the QPTs.

\begin{figure}
\leavevmode
   \epsfxsize=0.33\textwidth\epsfbox{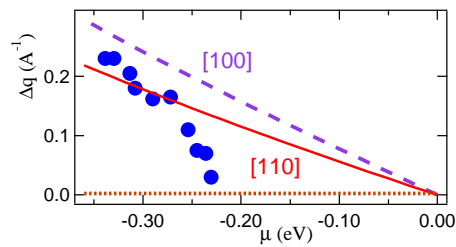}
\vskip0.5cm
\caption{Widths of susceptibility plateaus near $\vec Q$, as a function of 
chemical potential $\mu$.  Data from Ref~\protect\onlinecite{BalBo} (circles).}
\label{fig:6}
\end{figure}

Figure~\ref{fig:1} shows the bare susceptibility as a function of $\vec q$
both in and out of the hot spot regime.  At each doping in the hot spot
regime, there is a plateau in $\vec q$-space centered at $\vec Q$, where
$Re(\chi_{0q})$ is nearly constant, Fig.~\ref{fig:1}a.  
This plateau is highly {\it asymmetric} between electron and hole
doping: the width of this plateau $q_c$ shrinks to zero at $x_C$,
Fig.~\ref{fig:1}c, but actually has its largest value at $x_H$.  The
$q$-plateau half-width $q_c$ is the point where the $\vec Q+\vec q_c
$-shifted-FS no longer overlaps the original FS.  For displacements along
the $[110]$ direction, $q_c$ is found from
\begin{equation}
\sin{q_{cx}\over 2}=-{1\over\tau}-\sqrt{{1\over\tau^2}-{\mu\over 4t'}},
\label{eq:3}
\end{equation}
with $\tau =2t'/t$, while along the $[100]$ direction, $q_c=2sin^{-1}(-\mu /2t)
$.  The plateau is a Fermi surface caliper: Eq.~\ref{eq:3} represents the
distance in $\vec q$-space between the nodal point and the zone diagonal.
The cutoffs are plotted in Fig.~\ref{fig:6}.  The experimental 
data\cite{BalBo,Bou} will be discussed below.

For doping off of the plateaus, either for $x>x_H$, Fig.~\ref{fig:1}b, or
$x<x_c$, Fig.~\ref{fig:1}c, there are {\it two independent QPTs}.  First,
the magnetization vector changes from commensurate ($\vec q=\vec Q$) or
weakly incommensurate on the plateau to strongly incommensurate off of the
plateau, due to Kohn anomalies\cite{OPfeut,ETT}.  However, while this
involves a transition between two different forms of magnetic order, there
is a second transition.  Due to the rapid falloff of the {\it magnitude}
of $\chi_0$, there is a Slater-type transition to a nonmagnetic state.  It
is this latter transition which matches experimental observations for the
cuprate QPTs.  The falloff is so sudden that the strongly incommensurate
susceptibilities have not yet been seen.

Figure~\ref{fig:6} compares the theoretical plateau widths near the $H$-point 
with the {\it measured}\cite{BalBo,Bou} plateau widths in YBa$_2$Cu$_3$O$_{7-
\delta}$ in the normal state, along the [110]-direction.  [Note that the
broad feature analyzed in Ref.~\onlinecite{BalBo} is distinct from the
resonance peak.] For low
doping ($\mu >-0.28eV$) the widths are only weakly $\omega$-dependent, and
are measured in the low-$\omega$ limit.  For higher doping there is a spin
gap at the lowest frequencies, but since this may be a sign of a competing
instability, the magnetic linewidth can be estimated at the lowest
frequencies above the gap.  Particularly at high doping, the agreement
with the calculated plateau width is quite good: i.e., {\it neutron
scattering is measuring  the plateau width}, which is {\it not} indicative
of the correlation length.  The `deviations' near half filling ($\mu
=-0.2eV$) are in fact the expected SCR behavior, with the system
developing long-range N\'eel order, $\xi\sim 1/(\Delta
q)\rightarrow\infty$.  In turn, the large values of $q_c$ can explain why
$T_N\rightarrow 0$ at such a low hole-doping: the dotted line in
Fig.~\ref{fig:6} shows the criterion $\xi /a=100$, which is the threshold
for a finite Neel temperature in the electron-doped cuprates\cite{Gre}.

\begin{figure}
\leavevmode
   \epsfxsize=0.33\textwidth\epsfbox{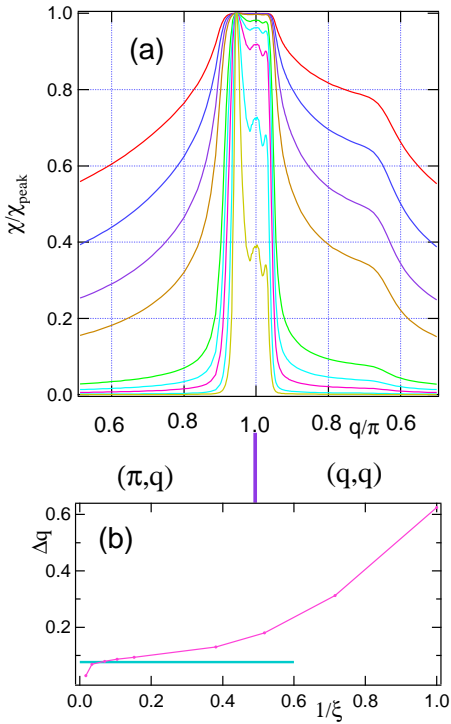}
\vskip0.5cm
\caption{(a) Normalized susceptibility as the QPT is approached, for 
$x=0.1$, $T=100K$.  From broadest to narrowest, the susceptibility
denominator is $\delta_0$ = 1, 0.51, 0.27, 0.145, 0.023, 0.011, 0.0048,
0.0011, and 0.00027. (b) Peak width $\Delta q$ vs
normalized correlation length $\xi =1/\sqrt{\delta_0}$.}
\label{fig:7}
\end{figure}

The relation between linewidth $\Delta q$ and correlation length $\xi$ can
be readily demonstrated.  In a self-consistent renormalization (SCR)
calculation\cite{MKII,Mor}, the RPA susceptibility is renormalized to 
$\chi_Q=\chi_{0Q}/\delta$, with $\delta =1-U\chi_{0Q}+\lambda$, and
$\lambda$ is a fluctuation-induced correction which keeps $\delta >0$ for
$T>0$.  For $\vec q=\vec Q+\vec q{}'$ close to $\vec Q$, 
\begin{equation}
\delta\propto\xi^{-2}+q'^2.
\label{eq:11}
\end{equation}
Thus if Eq,~\ref{eq:11} holds over a wide $q$-range, the peak width is a
measure of the magnetic correlation length.  In the presence of the
susceptibility plateau, this is no longer the case, Fig.~\ref{fig:7}.  
Figure~\ref{fig:7}a shows the SCR susceptibility (normalized to its peak
value) at a typical hole doping, for several values of $\delta_0=\delta
(q'=0)$.  In Fig.~\ref{fig:7}b, the resulting half-width $\Delta q$ is
plotted as a function of a normalized $\xi =1/\sqrt{\delta_0}$.  It can be
seen that for an extended range of $\xi$ the linewidth is unrelated to
$\xi$ and measures the plateau width, as expected from Fig.~\ref{fig:6}.
In the present calculation, the magnetization becomes incommensurate, 
causing $\Delta q$ to start decreasing again and $\rightarrow 0$ as
$\xi\rightarrow\infty$.  However, this is due to a giant amplification of
fine structure on the plateau close to the ordering transition; in reality
such fine structure may be washed out by fluctuations.

\begin{figure}
\leavevmode
   \epsfxsize=0.33\textwidth\epsfbox{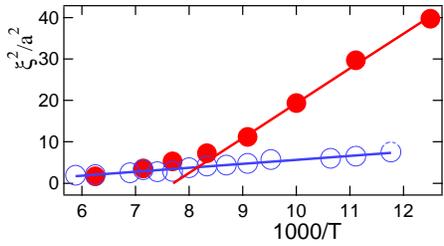}
\vskip0.5cm
\caption{Measured\protect\onlinecite{OBAM} magnetic coherence lengths in
YBCO, plotted according to Eq.~\protect\ref{eq:22}.  Full circles for
underdoped, open circles for optimally doped samples.  Straight lines are
fits to Eq.~\protect\ref{eq:22} with parameters $a$, $b$ = 8400K, 64.7
(960K, 3.93) for $T_c$ = 66K (90K).}
\label{fig:8}
\end{figure}

The presence of the susceptibility plateau can affect the $T$-dependence 
of $\xi$.  In two-dimensions, fluctuations prevent a finite temperature
Neel transition, but in the {\it renormalized classical} regime the
susceptibility is expected to diverge exponentially as $T\rightarrow 0$,
leading to a similar divergence of $\xi$.  However, the susceptibility
must satisfy a sum rule, and if the peak width stops decreasing (due to
the plateau), this sum rule can be saturated, leading to a weaker
temperature dependence for $\xi$.  The fluctuation-dissipation theorem can
be written as\cite{Mor,Ful}
\begin{equation}
<M^2>=-\int{d\omega\over\pi}n(\omega )\int{d^2q\over (2\pi
)^2}({c_x+c_y\over 2})Im\chi (\vec q,\omega )
\label{eq:21}
\end{equation}
where $<M^2>$ is the mean square local amplitude of nearest neighbor spin
fluctuations and $n$ is the Bose function.  The integral can be separated
into a part over the plateau and a background term from regions far
from\cite{FM} $\vec Q=(\pi ,\pi )$, and approximately evaluated as
$A/T-B=q_c^2/\delta_0$ where $q_c$ is the plateau width, or
\begin{equation}
\xi^2={a\over T}-b,
\label{eq:22}
\end{equation}
which holds at low temperatures, $T<<a/b$.  Figure~\ref{fig:8} plots
Eq.~\ref{eq:22} for the NMR derived coherence lengths\cite{OBAM} for 
optimally doped $T_c=90K$, and underdoped $T_c=66K$ YBa$_2$Cu$_3$O$_{7-
\delta}$ (YBCO).  It can be seen that both curves follow Eq.~\ref{eq:22}
at low $T$.  Hence, the plateaus explain the
weak divergence of the correlation length in hole doped cuprates.  (Note
that if there were no plateaus, $q_c\rightarrow 0$, there would be no
constraint on $\xi$ -- the width would decrease as the peak height
increased -- and the exponential divergence would be recovered, as in
the electron-doped cuprates.)  Finally, since $a$ is proportional to 
the square of the staggered magnetization, the decrease of the $a$
coefficient with doping may signal the proximity of the system to a
magnetic QPT.

Thus the present model can explain the high-T properties of the pseudogap
in the hole-doped cuprates.  At lower temperatures the hole-doped cuprates
are also susceptible to a number of competing instabilities, most notably
superconductivity, but also including stripe physics.  Certainly, the
broad $q$-plateau is conducive to incommensurate modulations, 
Fig.~\ref{fig:7}.  While these competing orders will complicate detailed
calculations, they are unlikely to significantly modify the picture of the
Mott gap collapse presented here.
%The magnetic susceptibility should still have the form of
%Eq~\ref{eq:22}, with the competing order mainly contributing to the
%$b$-coefficient.

In conclusion, the broad $q$-plateaus near the H-point can reconcile the
neutron scattering and NMR measurements of the correlation length -- NMR
should see a divergence, albeit weak $\xi\sim T^{-1/2}$, as $T\rightarrow
0$.  The neutron {\it linewidth} measurement should not see such a
divergence, but the divergence should be reflected in neutron measurements
as a $1/T$ growth of the peak intensity.  A resolution of the correlation
length problem along these lines would mean that the same Mott gap collapse
can explain the QPTs for both electron and hole doped cuprates -- indeed
the disappearence of hot spots provides a {\it natural phase boundary}
for QPTs. Superconductivity near an AFM or ferromagnetic QPT has recently
been observed in a number of systems\cite{MaLo}. 

{\bf Acknowledgments}: This work has been supported by the Spanish Secretaria de
Estado de Educaci\'on y Universidades under contract n$^o$ SAB2000-0034,
and by the U.S.D.O.E. Contract W-31-109-ENG-38 and has benefited from the
allocation of supercomputer time at the NERSC and the Northeastern
University Advanced Scientific Computation Center (NU-ASCC).  Part 
of the work was carried out while I was on sabbatical at the Instituto de
Ciencia de Materiales de Madrid (ICMM), CSIC, Cantoblanco, E-28049 Madrid,
Spain.  I thank Paco Guinea and Maria Vozmediano for inviting me and I
thank them and Martin Greven for many stimulating discussions.  


\begin{references}
\bibitem{Tdep}The alert reader might note that there is a strong
temperature dependence of $\chi_{0Q}$ near $x_H$ even below 100K.
\bibitem{FM}Near optimal hole doping, there are significant ferromagnetic
fluctuations, Fig.~\ref{fig:1}a (near $\Gamma$).
\end{references}
\end{document}